\documentclass[conference]{IEEEtran}
\IEEEoverridecommandlockouts

\usepackage{cite}
\usepackage{amsmath,amssymb,amsfonts}
\usepackage{graphicx}
\usepackage{textcomp}
\usepackage{xcolor}
\usepackage{url}
\usepackage[linesnumbered,ruled]{algorithm2e}

\usepackage[T1]{fontenc}
\usepackage[scaled=0.9]{inconsolata} 
\usepackage{listings}

\lstdefinestyle{finvet}{
  language=Python,
  basicstyle=\ttfamily\footnotesize,
  columns=fullflexible,
  breaklines=true,
  breakatwhitespace=true,
  keepspaces=true,
  showstringspaces=false,
  upquote=true,
  numbers=left,
  numberstyle=\tiny,
  numbersep=8pt,
  frame=single,
  rulecolor=\color{black},
  xleftmargin=1.5em,
  xrightmargin=1.5em,
  aboveskip=0.75\baselineskip,
  belowskip=0.75\baselineskip,
  captionpos=b
}

\def\BibTeX{{\rm B\kern-.05em{\sc i\kern-.025em b}\kern-.08em
    T\kern-.1667em\lower.7ex\hbox{E}\kern-.125emX}}

\title{FinVet: A Collaborative Framework of RAG and External Fact-Checking Agents for Financial Misinformation Detection}

\author{\IEEEauthorblockN{Daniel Berhane Araya, Duoduo Liao}
\IEEEauthorblockA{\textit{College of Engineering and Computing} \\
\textit{George Mason University}\\
Fairfax, VA, USA \\
\{dberhan4, dliao2\}@gmu.edu}
}

\begin{document}

\maketitle

\begin{abstract}
Financial markets face growing threats from misinformation that can trigger billions of dollars in losses within minutes. Most existing approaches lack transparency in their decision-making and provide limited attribution to credible sources. We introduce FinVet, a novel multi-agent framework that integrates two Retrieval-Augmented Generation (RAG) pipelines with external fact-checking through a confidence-weighted voting mechanism. FinVet employs adaptive three-tier processing that dynamically adjusts verification strategies based on retrieval confidence, from direct metadata extraction to hybrid reasoning to full model-based analysis. Unlike existing methods, FinVet provides evidence-backed verdicts, source attribution, confidence scores, and explicit uncertainty flags when evidence is insufficient. Experimental evaluation on the FinFact dataset shows that FinVet achieves an F1 score of 0.85, which is a 10.4\% improvement over the best individual pipeline (Fact-Check pipeline) and 37\% improvement over standalone RAG approaches.

\end{abstract}

\begin{IEEEkeywords}
Financial misinformation detection, Retrieval-Augmented Generation (RAG), fact-checking, large language models (LLMs), confidence-based voting, explainable AI, Google Fact Check API
\end{IEEEkeywords}

\section{Introduction}
Financial markets are increasingly susceptible to misinformation, in which a single false claim can trigger billions of dollars in losses within minutes. 
For instance, a 2013 tweet from a hacked Associated Press account falsely reporting a White House explosion “wiped out \$130 billion from the stock market" \cite{Qian2018}. More recently, in May 2023, a viral fake image of a Pentagon explosion caused the Dow Jones Industrial Average to drop 80 basis points (bps) and the S\&P 500 to fall by 26 bps \cite{Parsan2024}. These incidents underscore the urgent need for a robust and transparent framework capable of detecting and mitigating financial misinformation.
Financial misinformation is defined as false or misleading information related to economic or financial matters that can influence investor decisions, market behavior, or regulatory outcomes \cite{Rangapur2024}.  It can also include fraudulent schemes aimed at defrauding individuals or institutions.

The advent of generative AI has triggered a wave of AI-driven tools, escalating misinformation, disrupting financial markets, and fueling a surge in fraud. Malicious actors now exploit generative AI to create deepfake videos, synthetic voices, and fraudulent documents—enabling large-scale deception that challenges financial institutions' detection capabilities.
Social media platforms, central to the rapid dissemination of misinformation, exacerbate this growing threat. According to Deloitte \cite{Deloitte2023}, AI-enabled fraud losses in the United States are projected to increase from \$12.3 billion in 2023 to \$40 billion by 2027.

Identifying and mitigating financial misinformation is crucial for financial institutions, investors, and regulators. Left unchecked, misinformation can manipulate stock prices, mislead investors, and destabilize markets. To this end, this paper introduces FinVet, a novel multi-agent framework designed to address the aforementioned challenges through the integration of two Retrieval-Augmented Generation (RAG) \cite{lewis2020} pipelines and a Fact-Check Pipeline. The RAG components utilize an external knowledge source, such as a vector store, which can be tailored to any domain-specific dataset. The Fact-Check Pipeline combines an external fact-checking source with a fallback mechanism leveraging a Large Language Model (LLM) when direct evidence is unavailable. The outputs of these independent pipelines are integrated via a confidence-based voting mechanism to deliver a final verdict. This integrated approach provides a robust and adaptable misinformation detection system, offering improved accuracy and explainability. In addition to flagging misinformation, FinVet presents evidence-backed justifications, clearly cites source origins, and assigns a confidence score quantifying the system’s certainty in its verdict.

The main contributions of this paper are:

\begin{enumerate}
    \item A novel multi-agent verification framework that systematically integrates dual RAG pipelines with external fact-checking orchestrated using a confidence-weighted voting mechanism, addressing the limitations of single-pipeline approaches in financial misinformation detection.
    
    \item An adaptive three-tier processing strategy that dynamically selects verification approaches based on retrieval confidence scores: (i) direct metadata extraction for high-confidence retrievals, (ii) hybrid context-model reasoning for moderate confidence, and (iii) pure model-based analysis for low-confidence scenarios, optimizing both computational efficiency and verification accuracy.

    \item A hierarchical confidence-weighted integration mechanism that combines heterogeneous verification outputs to provide a more robust verdict.
    
    \item Comprehensive empirical validation demonstrating that this collaborative approach achieves an F1 score of 0.85, which is a 10.4\% improvement over the best individual pipeline.
    
\end{enumerate}

The remaining sections of the paper are organized as follows: Section II reviews related work; Section III provides a summary of the contributions; Section IV describes the proposed framework; Section V presents the experimental results and discussion; Section VI discusses ethical considerations; and Section VII concludes the paper.

\section{Related Work}
Financial misinformation detection is a complex task that has been extensively researched using various techniques. Although most existing approaches predominantly rely on deep learning-based methods, recent research has increasingly explored Large Language Models (LLMs) for this purpose. In this paper, based on the objectives and methodologies of the reviewed studies, we categorize financial misinformation detection research into deep learning-based and LLM-based techniques which reflect the evolving landscape of approaches in this domain.

\subsection{Deep Learning-Based Techniques}

Several previous studies utilize machine learning and deep learning techniques, for financial misinformation detection.

Dmonte \textit{et al.} \cite{dmonte2025gmu} instruction-tune \mbox{LLaMA-3.1-8B} using claim–justification pairs and also explore a prompting-based few-shot approach. Similarly, Purbey \textit{et al.} \cite{purbey2025sharedtasks} use a sequential fine-tuning strategy: first, training LLMs for classification, then for joint explanation generation to optimize end-to-end performance on the financial misinformation detection task.

Zhang and Liu \cite{Zhang2023} focus on identifying financial news in the Chinese market with a deep learning approach. They propose a dual sub-network model, employing BERT-Chinese-wwm embeddings \cite{Devlin2018},\cite{cui2021} to generate rich word representations, which feed into three parallel Convolutional Neural Networks (CNNs) with varying kernel sizes to extract multi-granular semantic features from article content. 

Kamal, Mohankumar, and Singh\cite{Kamal2023} tackle financial misinformation with their Fin-MisID model. They leverage RoBERTa \cite{Liu2019} for contextual representations, feeding them into multi-channel networks of parallel Convolutional Neural Networks (CNNs) with varying kernel sizes, Bidirectional Gated Recurrent Units (BiGRU) for sequential patterns, and attention mechanisms to emphasize critical tokens.

Nasir, Khan, and Varlamis\cite{Nasir2021} propose a hybrid deep learning architecture combining CNN and Long Short-Term Memory networks (LSTMs) for fake news detection. CNNs are used to extract local textual patterns and n-gram features, which are then passed to LSTMs to capture long-range dependencies in the sequence. The model is trained end-to-end, enabling joint learning of spatial and temporal features. This approach leverages the complementary strengths of CNNs and RNNs for improved representation of news content.

X. Zhang, Q. Du, and Z. Zhang\cite{zhang2022} design a theory-driven machine learning system for financial disinformation detection, grounded in Truth-Default Theory (TDT). Their framework operationalizes five key deception cues—communication context and motive, sender demeanor, third-party information, coherence, and correspondence—into quantifiable features. These features are extracted from financial news articles, propagation patterns across platforms, author metadata, and firm-level financial indicators. The system integrates these heterogeneous signals into a supervised learning pipeline, offering interpretable and domain-sensitive disinformation detection.

Zhi et al.\cite{zhi2021financial} propose a multi-fact CNN-LSTM model that integrates multiple dimensions of information for financial fake news detection. The model combines news content, market data, user comments, and source credibility into a unified framework. Character-level CNNs are employed to extract granular textual features, which are then encoded through LSTM layers to capture temporal dependencies. An attention mechanism models the interaction between news content and user comments, and all components are fused through weighted aggregation for the final prediction.

The deep learning-based approaches described above primarily focus on classification-oriented architectures that rely on features extracted from news text and propagation patterns. Although these methods demonstrate strong performance in identifying financial misinformation, they typically operate with limited transparency in their decision-making process. In contrast, FinVet introduces a novel multi-agent framework that combines RAG pipelines with an external fact-checking system, orchestrated through a confidence-based voting mechanism. Rather than relying solely on internal embeddings or latent features, FinVet retrieves verifiable evidence from a domain-specific vector store and external sources, providing rationale-backed predictions with explicit confidence scores and citation of information sources.

Furthermore, FinVet employs a tiered verification strategy: highly relevant retrieved evidence leads to direct labeling; moderately relevant evidence is combined with LLM reasoning; and in the absence of external context, the system relies on structured, role-based prompts to elicit reasoned judgments from LLMs. This adaptive methodology not only improves classification performance but also improves explainability.

\subsection{LLM-based Approaches}

Large Language Models (LLMs) have recently been explored for misinformation detection. Leite, Razuvayevskaya, Bontcheva, and Scarton \cite{leite2024} analyze whether LLMs can generate weak labels based on 18 credibility signals, which are then aggregated using weak supervision to predict content veracity. Their approach demonstrates that zero-shot LLM-based credibility labeling, combined with signal aggregation, can outperform traditional classifiers without requiring ground-truth labels. 

Liu et al.\cite{Liu2024} developed an instruction-tuning dataset, the Financial Misinformation Detection Instruction Dataset (FMDID), by combining two existing datasets: FinFact \cite{Rangapur2024} and FinGuard \cite{Martin2021}. Using their developed dataset, the authors fine-tuned LLaMa2-chat-7b \cite{Touvron2023} and Llama-3.1-8B-Instruct \footnote {https://huggingface.co/meta-llama/Llama-3.1-8B-Instruct}. They compared their model's performance against baseline models, including Pre-trained Language Models (PLMs) like BERT and RoBERTa, as well as LLMs such as LLaMA2, LLaMA3.1, and ChatGPT \cite{OpenAI2024}.

Wan et al. \cite{wan2024} propose DELL, a multi-stage misinformation detection framework that strategically integrates LLMs to improve veracity assessment. DELL begins by generating synthetic user-news interaction networks through various LLM-generated news reactions. It uses six explainable proxy tasks, such as sentiment, stance, and framing to enrich contextual understanding with LLM-generated explanations. Finally, it employs an LLM-based expert ensemble, where task-specific predictions and confidence scores are selectively aggregated to produce calibrated, rationale-backed misinformation judgments.

In contrast to the aforementioned research approaches, FinVet integrates two RAG pipelines with an online fact-checking tool, using a confidence score-based voting mechanism, ensuring claims are validated using multiple pathways. Moreover, by leveraging a domain-specific knowledge base, FinVet's architecture can easily adapt to emerging financial misinformation patterns while maintaining decision transparency that is crucial for financial stakeholders.

To the best of our knowledge, no prior work in financial misinformation detection has combined dual RAG pipelines, external fact-checking services, and confidence-based verdict integration within a unified, transparent architecture.

\section{Summary of Contributions}

This paper presents FinVet, a multi-agent verification framework that integrates evidence retrieval, external fact-checking, and confidence-based reasoning into a unified decision-making architecture. By directly addressing the limitations of existing approaches, particularly their reliance on single verification pipelines and limited transparency, FinVet delivers interpretable, adaptive, and evidence-based results tailored for financial misinformation detection. The key contributions of this study are summarized below:

\begin{itemize}
    \item \textit{Multi-Agent Verification Architecture:} FinVet employs three independently operating verification pipelines—two RAG pipelines and a Fact-Check Pipeline that integrates an external fact-checking source with an LLM fallback. This ensemble design enhances redundancy, promotes model diversity, and improves verification reliability.

    \item \textit{Three-Tier Similarity-Based Evidence Processing:} A cosine-similarity–driven mechanism dynamically adjusts the reasoning pathway based on the relevance of retrieved evidence. High similarity prompts direct evidence extraction, moderate similarity invokes hybrid model reasoning, and low similarity triggers expert-mimicking role-based analysis. 

    \item \textit{Confidence-Based Verdict Integration:} A hierarchical aggregation mechanism combines outputs from all verification pipelines and weights them by confidence, ensuring that high-certainty responses are prioritized in the final verdict.

    \item \textit{Transparent Verdicts with Source Attribution:} FinVet outputs not only claim labels but also supporting evidence sources and confidence scores, enhancing interpretability and aligning with ethical standards for decision-making in high-stakes financial domains.

    \item \textit{Empirical Validation through Baseline and Ablation Studies:} FinVet was evaluated on the FinFact dataset against multiple baselines, including zero-shot LLMs, chain-of-thought prompting, RAG-only models, and standalone fact-checking systems. It achieved an F1 score of 0.85, outperforming all baselines by 8–36\%. Ablation experiments further confirm the additive contribution of each verification component.

    \item \textit{First Integrated Application in Financial Misinformation Detection:} To our knowledge, this is the first work to combine RAG, external fact-checking, and confidence-based voting within a unified system for financial misinformation detection.
\end{itemize}

These contributions collectively position FinVet as a scalable, interpretable, and empirically validated framework that advances the state of the art in financial claim verification.

\section{Methodology}
This research proposes FinVet shown in Figure 1, and elaborated in Algorithm 1. FinVet is a novel framework that employs a multi-pipeline architecture for financial claim verification. The framework consists of four primary components: (A) Data Processing and Vector Storage, (B) Claim Verification Pipelines, (C) Results Normalization, and (D) Verdict Integration and Reporting. These components are described in the following sections.

\begin{figure*}[t]
  \centering
  \includegraphics[width=0.9\textwidth]{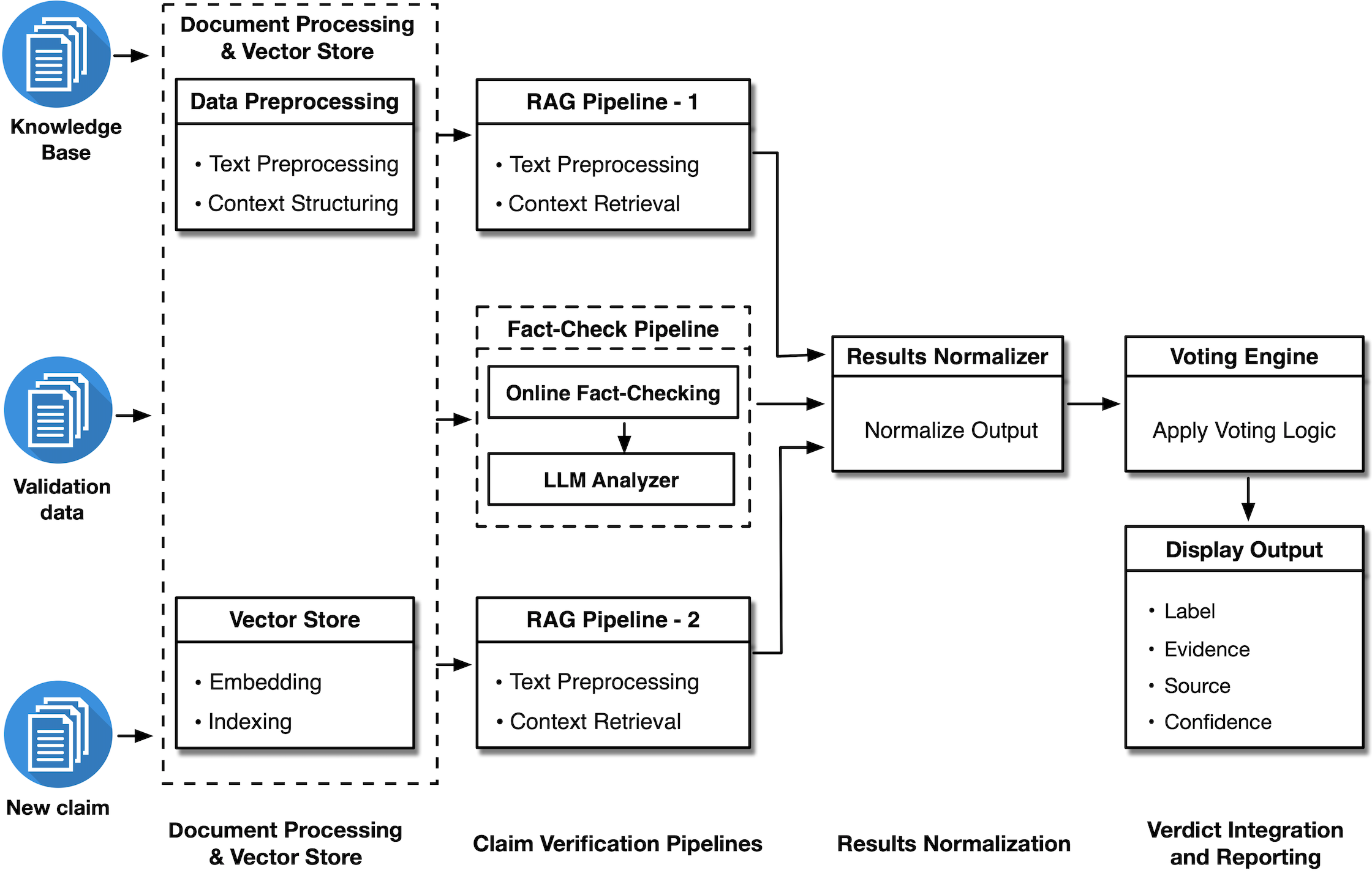} 
  \caption{FinVet: A Collaborative Framework of RAG and External Fact-Checking Agents for Financial Misinformation Detection}
  \label{fig:experiments}
\end{figure*}

\begin{algorithm}[t!]
\SetKwInOut{Input}{Input}
\SetKwInOut{Output}{Output}
\Input{Financial claim $C$, Vector store $\mathcal{V}$, External fact-check API $\mathcal{F}$, Similarity thresholds $\theta_{high} = 0.6$, $\theta_{med} = 0.4$}
\Output{Verdict $\mathcal{L}$ (true/false/NEI), Evidence $\mathcal{E}$, Source $\mathcal{S}$, Confidence $\rho$}
\BlankLine
\tcc{RAG Pipeline Processing}
\For{$i \gets 1$ \textbf{to} $2$}{
    $R_i \gets$ \texttt{RAGPipeline}($C$, $\mathcal{V}$, $M_i$)\;
    \tcc{$M_1$: LLaMA-3.3-70B, $M_2$: Mixtral-8x7B}
}
\BlankLine
\tcc{Fact-Check Pipeline Processing}
$R_3 \gets$ \texttt{FactCheckPipeline}($C$, $\mathcal{F}$)\;
\BlankLine
\tcc{Results Normalization}
\For{$j \gets 1$ \textbf{to} $3$}{
    $\hat{R}_j \gets$ \texttt{NormalizeResult}($R_j$)\;
}
\BlankLine
\tcc{Verdict Integration}
\eIf{$R_3$.source = ``External Fact-Check''}{
    \textbf{return} $R_3$\;
}{
    $k^* \gets \arg\max_k \hat{R}_k$.confidence\;
    \eIf{$\forall k : \hat{R}_k$.confidence = 0}{
        \textbf{return} \{label: ``NEI'', evidence: ``Insufficient information'',\\
        \hspace{3.5em} source: ``No evidence'', confidence: 0\}\;
    }{
        \textbf{return} $\hat{R}_{k^*}$\;
    }
}
\caption{FinVet: Financial Misinformation Detection Framework}
\label{alg:finvet}
\end{algorithm}

\subsection{Data Processing and Vector Store}

This component serves as FinVet's knowledge base by transforming financial information into semantically rich, retrievable representations. First, financial texts are decomposed into contextually meaningful claim–evidence pairs, preserving critical metadata relationships. This approach balances granularity and context, thus improving retrieval precision. Second, each text segment is embedded into a high-dimensional vector space using semantic encodings that capture domain-specific relationships. This enables nuanced concept-driven evidence retrieval. Finally, the origin of each piece of evidence is tracked by storing associated metadata throughout the verification pipeline, ensuring that FinVet’s outputs are grounded in traceable source-backed justifications.

\subsection {Claim Verification Pipelines}

The Claim Verification Pipelines represent the core analytical engines of FinVet, and it is comprised of two complementary verification approaches: two RAG Verification Pipelines and a Fact-Check Pipeline. These pipelines process financial claims through different methodologies, each bringing unique analytical strengths to determine the veracity of a claim. Using multiple verification strategies based on different approaches, the system attempts to achieve a more robust and nuanced assessment than any single approach could provide.

\subsubsection{RAG Verification Pipelines}
The RAG Verification Pipelines implement a retrieval-augmented generation approach that grounds verification in contextual knowledge. Each pipeline retrieves relevant financial information from the vector store and uses this context in its verification process. FinVet employs two distinct RAG pipelines that are implemented using different underlying language models. The framework is inherently extensible, allowing for the integration of additional RAG pipelines to further enhance the verification process. This modular design enables the incorporation of specialized verification pathways tailored to particular financial domains or claim types, such as market predictions, earnings reports, or macroeconomic indicators. Moreover, while the current RAG implementations use a shared vector store, the architecture supports the integration of multiple domain-specific knowledge bases that can be used for different financial sectors, regulatory frameworks, or historical time periods, enabling the system to triangulate evidence using multiple sources.

Both RAG pipelines follow a similar workflow: preprocessing the claim text, retrieving relevant context from the vector store based on semantic similarity, and generating structured verification outputs including label determination (true/false/Not Enough Information (NEI)), evidence generation, confidence scoring, and source attribution. The dual pipeline approach helps mitigate individual model biases through model diversity.

Central to FinVet's RAG pipeline is the three-tier processing step that dynamically adapts the verification process based on retrieval confidence scores and it is described in Algorithm 2. The system determines the relevance of the evidence using cosine similarity. This adaptive approach works as follows:

\begin{itemize}
    \item \textit{Case 1}: When highly relevant evidence is found (similarity score $>=$ 0.6), the system extracts the label, evidence, and source directly from the retrieved metadata without additional LLM reasoning. The evidence is taken directly from the retrieved sentence and the source attribution is preserved by directly passing through the original source from the metadata's records. The confidence score reflects the similarity score, and higher values represent closely matching evidence;
    
    \item \textit{Case 2}: For moderately relevant evidence (0.4 $<=$ Similarity Score $<$ 0.6), a hybrid approach is used by combining the retrieved context with model reasoning, providing both the retrieved information and claim to the foundation model with explicit instructions to evaluate the claim using the provided context. The evidence is generated by the foundation model and the source attribution maintains the original retrieved sources when the model validates their relevance; if the model determines the retrieved information is insufficient, it indicates "Parametric Knowledge" as the source. Confidence is calculated as the average of the retrieval similarity and the model's self-reported confidence, balancing evidence relevance with model certainty;
    
    \item \textit{Case 3}: When no strongly relevant evidence is available (similarity $<$ 0.4), the system falls back to a role-based prompting that instructs the foundation model to analyze the claim from multiple expert perspectives (Financial Analyst, Political Misinformation Specialist, Government Policy Analyst, and Investigative Journalist). In this tier, evidence is entirely generated by the foundation model's reasoning process and explicitly identified as model-generated content. Source attribution is consistently marked as "Parametric Knowledge" to maintain transparent distinction between retrieved and generated content. Confidence is based entirely on calibrated assessment of the foundation model.
\end{itemize}

The similarity cut-off values used in the above three-tier approach were obtained from multiple rounds of tests that were performed using the FinFact dataset.

This adaptive processing framework optimizes both computational efficiency as well as cost, by invoking increasingly complex verification only when necessary.

\subsubsection{Fact-Check Pipeline}

The Fact-Check Pipeline provides a modular verification layer that complements the RAG-based methodology by linking FinVet to externally verified, authoritative information sources. It can interface with any structured fact-checking repository, whether academic, governmental, or institutional. The pipeline operates through two distinct stages:

\begin{itemize}
    \item \textit{External Verification Component}: 
    This component queries externally maintained fact-checking repositories to determine whether the input claim has already been verified. If a matching record exists, the system retrieves the publisher’s verdict, explanatory rationale, and source reference, then maps these elements into FinVet’s standardized output format comprising verdict, evidence, and source. Verified claims obtained from external repositories are treated as authoritative and assigned a fixed confidence value of 1.0. In practice, this component can connect to publicly available repositories such as Google Fact Check Tools, PolitiFact, Snopes, or Reuters Fact Check, while the architecture also supports integration with other domain-specific fact-checking APIs.
    
    \item \textit{LLM Analyzer}: 
    When no verified record is found externally, the system employs role-based prompting to an LLM to analyze the claim and generate verdict and supporting evidence. The source of the verdict from this stage is explicitly labeled 'Parametric Knowledge', distinguishing model-generated reasoning from externally verified information. The confidence value is also derived from the LLM’s internal scoring or self-assessed certainty estimate.
\end{itemize}

To limit potential error propagation, the prompting template constrains the model to cite only information derived from the input claim, to flag ambiguity or insufficient context with a “Not Enough Information (NEI)” label instead of producing unsupported conclusions, and to perform basic internal consistency checks such as temporal and numerical validation. These measures ensure that the reasoning process remains bounded, transparent, and traceable within FinVet’s verification workflow.

\subsection{Results Normalization}
At this stage, FinVet has the verification results of a claim using three independent pipelines: two RAG pipelines and an external Fact-Check Pipeline. This component transforms these diverse outputs into a consistent format. It normalizes labels to lower case ("true", "false" or "nei") and standardizes confidence score formatting which is crucial for the next and final component. 
\begin{algorithm}[t!]
\SetKwInOut{Input}{Input}
\SetKwInOut{Output}{Output}
\Input{Claim $C$, Retrieved documents $\{d_1, ..., d_k\}$, Similarity scores $\{s_1, ..., s_k\}$, Thresholds $\theta_{high} = 0.6$, $\theta_{med} = 0.4$, Model $M$, Expert roles $\mathcal{R}$}
\Output{Result $R$ = \{label, evidence, source, confidence\}}
\BlankLine
\tcc{Check retrieved documents}
\eIf{$k = 0$}{
    \tcc{No documents retrieved}
    result $\gets$ RoleBasedAnalysis($M$, $C$, $\mathcal{R}$)\;
    label $\gets$ result.label\;
    evidence $\gets$ result.evidence\;
    source $\gets$ ``Parametric Knowledge''\;
    confidence $\gets$ ModelConfidence($M$)\;
}{
    \tcc{Evaluate Similarity}
    $s_{max} \gets \max\{s_1, ..., s_k\}$\;
    $d_{best} \gets$ document with $s_{max}$\;
    \BlankLine
    
    \eIf{$s_{max} > \theta_{high}$}{
        \tcc{Case 1}
        label $\gets$ ExtractLabel($d_{best}$)\;
        evidence $\gets$ ExtractEvidence($d_{best}$)\;
        source $\gets$ ExtractSource($d_{best}$.metadata)\;
        confidence $\gets s_{max}$\;
    }{
        \eIf{$s_{max} \geq \theta_{med}$}{
            \tcc{Case 2}
            context $\gets$ CombineDocuments($\{d_1, ..., d_k\}$)\;
            result $\gets$ ModelReasoning($M$, $C$, context)\;
            label $\gets$ result.label\;
            evidence $\gets$ result.evidence\;
            \eIf{ModelValidatesRetrieved(result)}{
                source $\gets$ ExtractSource($d_{best}$.metadata)\;
            }{
                source $\gets$ ``Parametric Knowledge''\;
            }
            confidence $\gets$ Average($s_{max}$, result.confidence)\;
        }{
            \tcc{Case 3}
            result $\gets$ RoleBasedAnalysis($M$, $C$, $\mathcal{R}$)\;
            label $\gets$ result.label\;
            evidence $\gets$ result.evidence\;
            source $\gets$ ``Parametric Knowledge''\;
            confidence $\gets$ ModelConfidence($M$)\;
        }
    }
}
\BlankLine
\textbf{return} $R$ = \{label, evidence, source, confidence\}\;
\caption{RAG Processing Pipeline}
\label{alg:rag-processing}
\end{algorithm}
\subsection{Verdict Integration and Reporting}
This component combines the verdict of the above
three pipelines through a confidence-based voting mechanism. The voting process logic is as follows:

\begin{enumerate}
    \item If the Fact-Check Pipeline returns results from an external fact-checked source, these are automatically prioritized as the final verdict.

    \item  In typical cases, the system selects the result with the highest confidence score across all pipelines.
    
    \item If all pipelines return zero confidence scores, the system defaults to a "NEI" verdict, explicitly acknowledging insufficient evidence rather than making an unwarranted classification.
\end{enumerate}

Following this decision process, the reporting component generates an output that includes: the final verification label (true/false/nei), supporting evidence, source and confidence-score.

\section{Experimental Results and Discussion}
This section describes the experimental implementation and evaluation of FinVet. It includes the models and tools used and is followed by a comprehensive evaluation of its performance.

\subsection{Experimental Setup}
This subsection describes the technical components used to implement FinVet, including vector embedding and storage tools, LLMs, and external verification tools:

\subsubsection{Vector Embedding and Storage} 

FinVet was implemented using the text embedding model all-MiniLM-L6-v2 \cite{wang2020}\cite{reimers2019}. This sentence transformer model generates compact 384-dimensional embeddings that effectively capture semantic relationships between the concepts of a claim. The Facebook AI Similarity Search (FAISS) library with InVerted File with Flat storage (IVFFlat) architecture \cite{jegou2022} was used to index and retrieve the embeddings, enabling fast similarity search in large volumes of financial statements. 

\subsubsection{RAG Pipeline Implementation} The dual RAG pipelines were implemented using the following LLMs:
\begin{itemize}
    \item \textit{RAG Pipeline-1} was implemented using the LLaMA-3.3-70B-Instruct model\footnote{\url{https://huggingface.co/meta-llama/Llama-3.3-70B-Instruct}} \cite{grattafiori2024}, accessed through the Hugging Face API. This transformer-based model was selected for its strong general knowledge and reasoning capabilities.

   \item \textit{RAG Pipeline-2} was implemented using the Mixtral-8x7B-Instruct-v0.1 model\footnote{\url{https://huggingface.co/mistralai/Mixtral-8x7B-Instruct-v0.1}} \cite{jiang2024}, accessed through the Hugging Face API. This model was selected for its complementary reasoning patterns through its Mixture-Of-Experts (MOE) architecture.
\end{itemize}

\subsubsection{Fact-Check Pipeline} This pipeline was implemented as follows:
\begin{itemize}
    \item The external verification component was interfaced with the Google Fact Check Tools API \cite{googlefactcheck2025}.
    
    \item The LLM Analyzer component used the LLaMA-3.3-70B-Instruct model. 

\end{itemize}

The LLMs in this study were configured with a temperature of 0.10 to ensure a more deterministic output and max\_new\_tokens was set to 300 to balance response completeness with computational efficiency.

\subsection{Dataset and Evaluation Methodology}
FinVet was evaluated using the FinFact dataset \cite{Rangapur2024}, a comprehensive collection of financial claims with expert-verified labels and supporting evidence.

\begin{table}[htbp]
\caption{Label Distribution in the FinFact Dataset}
\centering
\small
\renewcommand{\arraystretch}{1.1}
\begin{tabular}{|l|c|c|}
\hline
\textbf{Label} & \textbf{Samples} & \textbf{Percentage} \\ \hline
False & 1,492 & 44.29\% \\ \hline
True & 1,275 & 37.85\% \\ \hline
Not Enough Information (NEI) & 602 & 17.87\% \\ \hline
\textbf{Total} & \textbf{3,369} & \textbf{100.00\%} \\ \hline
\end{tabular}
\label{tab:class_distribution}
\end{table}

The dataset exhibits a moderate class imbalance, as summarized in Table~\ref{tab:class_distribution}. To preserve representative class proportions across training and testing phases, stratified sampling was applied, allocating 85\% of the data for training and 15\% for testing. The training subset was utilized to populate the vector store, whereas the test subset was reserved exclusively for performance evaluation. Model performance was assessed using standard metrics, including accuracy (Acc.), precision (Prec.), recall (Rec.), and F1-score (F1). To address the effect of class imbalance, the F1-score was adopted as the primary evaluation metric in all comparative analyses.
\subsection{Baseline Evaluation}

FinVet is evaluated against three baseline approaches: RAG (LLaMA-3.3-70B-Instruct), and Zero-shot and Chain-of-Thought (GPT-3.5-turbo). Table~\ref{tab:baseline} shows the results. FinVet achieves an F1 score of 0.85, demonstrating substantial improvements over all baselines. The retrieval-augmented baseline, RAG (LLaMA-3.3-70B), achieves an F1 of 0.62, with notably higher precision (0.69) than recall (0.56), suggesting conservative prediction behavior when using retrieval alone. FinVet improves upon this by 37\% in F1 score while achieving more balanced precision (0.86) and recall (0.84).

The parametric baselines show further performance degradation. Zero-shot GPT-3.5 achieves an F1 of 0.51, with FinVet outperforming it by 67\%. Surprisingly, Chain-of-Thought prompting performs worse, dropping to 0.45 F1 with the lowest accuracy (0.37) despite moderate precision (0.56), indicating that explicit reasoning without grounded evidence may introduce errors in financial claim verification. FinVet's 89\% improvement over CoT underscores the critical importance of combining retrieval with external verification for this domain.

\subsection{Ablation Study}

To understand the contribution of individual components within the FinVet architecture, we conducted a systematic ablation study evaluating five configurations:

\begin{itemize}
    \item \textit{FinVet (Full System)}: The complete pipeline integrating two RAG modules, a Fact-Check pipeline, and confidence-weighted verdict integration.
    
    \item \textit{Fact-Check Pipeline}: A hybrid system combining Google Fact Check API with LLaMA-3.3-70B for fallback reasoning when there is no match in external fact-checks. This configuration excludes RAG pipelines.
    
    \item \textit{Google Fact Check Only}: Standalone external fact-checking API without any LLM support or RAG pipelines
    
    \item \textit{RAG (LLaMA-3.3-70B)}: Single RAG pipeline using LLaMA-3.3-70B, excluding fact-checking components.
    
    \item \textit{RAG (Mixtral-8x7B)}: Single RAG pipeline using Mixtral-8x7B, excluding fact-checking components.
\end{itemize}

Table~\ref{tab:ablation} shows the ablation results. The individual RAG pipelines achieve F1 scores of 0.49 (Mixtral-8x7B) and 0.62 (LLaMA-3.3-70B), providing moderate baseline performance. The Google Fact Check API alone also achieves an F1 score of 0.62. Combining the external fact checking API with LLM reasoning in the Fact-Check Pipeline improves performance to 0.77. Finally, FinVet's full integration of both RAG pipelines and fact-checking achieves an F1 score of 0.85, which is a 10.4\% relative improvement over the best individual component (Fact-Check Pipeline) and a 37\% improvement over the standalone RAG approaches. These empirical results demonstrate that, for financial claim verification, the proposed hybrid architecture integrating dual RAG pipelines with fact-checking capabilities significantly outperform each constituent verification method when deployed independently.

\begin{table}[htbp]
\caption{Performance Comparison of FinVet and Baseline Approaches}
\centering
\small 
\renewcommand{\arraystretch}{1.1}
\begin{tabular}{|l|c|c|c|c|}
\hline
\textbf{Approach} & \textbf{Acc.} & \textbf{Prec.} & \textbf{Rec.} & \textbf{F1} \\
\hline
FinVet & 0.84 & 0.86 & 0.84 & 0.85 \\
\hline
RAG (LLaMA-3.3-70B) & 0.58 & 0.69 & 0.56 & 0.62 \\
\hline
Zero-shot (GPT-3.5) & 0.50 & 0.52 & 0.50 & 0.51 \\
\hline
CoT$^{\mathrm{a}}$ (GPT-3.5) & 0.37 & 0.56 & 0.37 & 0.45 \\
\hline
\multicolumn{5}{l}{\scriptsize $^{\mathrm{a}}$CoT: Chain-of-Thought prompting.} \\
\end{tabular}
\label{tab:baseline}
\end{table}

\begin{table}[htbp]
\caption{Ablation Study of FinVet Components}
\centering
\small 
\renewcommand{\arraystretch}{1.1}
\begin{tabular}{|l|c|c|c|c|}
\hline
\textbf{Approach} & \textbf{Acc.} & \textbf{Prec.} & \textbf{Rec.} & \textbf{F1} \\
\hline
FinVet & 0.84 & 0.86 & 0.84 & 0.85 \\
\hline
Fact-Check Pipeline$^{\mathrm{a}}$ & 0.76 & 0.78 & 0.76 & 0.77 \\
\hline
Google Fact Check Only & 0.49 & 0.84 & 0.49 & 0.62 \\
\hline
RAG (LLaMA-3.3-70B) & 0.58 & 0.69 & 0.56 & 0.62 \\
\hline
RAG (Mixtral-8x7B) & 0.44 & 0.53 & 0.46 & 0.49 \\
\hline
\multicolumn{5}{l}{\scriptsize $^{\mathrm{a}}$Google Fact Check API + LLaMA-3.3-70B.} \\
\end{tabular}
\label{tab:ablation}
\end{table}

\section{Ethical Considerations}

Although FinVet is designed to enhance transparency in financial misinformation detection, its implementation raises several important ethical considerations.

\begin{enumerate}
\item \textit{Algorithmic and Data Bias}: The LLM models used in FinVet are subject to inherent algorithmic biases arising from their pretraining objectives and training data. These models often reflect and amplify societal, institutional, or geographic biases present in the large-scale corpora on which they are trained. As a result, the reasoning and evidence generation steps in FinVet may inherit these biases, potentially affecting the objectivity and fairness of its outputs.
The framework’s performance depends on the quality and diversity of its knowledge base. If training data contains biases or lacks representation of certain financial domains, FinVet may perform inconsistently across different types of financial claims. Lastly, the reliance on external fact-checking APIs introduces potential biases embedded in these services.
\item \textit{False Positives and Negatives}: No detection system is perfect. False positives could incorrectly flag legitimate financial information, potentially harming honest actors. In contrast, false negatives could allow harmful misinformation to spread. In this research, uncertainty handling has been implemented through explicit "NEI" verdicts to mitigate unwarranted classifications, but users should be aware of these limitations.
To mitigate these risks, transparency is advocated in the decision-making process, including clear documentation of pipeline biases and confidence thresholds. For future deployments, the incorporation of diverse sources of fact-checking and regular audits is recommended to ensure fairness and accountability. Interdisciplinary collaboration with ethicists and financial regulators is also encouraged to align the system with societal values and legal standards.
\end{enumerate}

\section {Conclusion}

This research introduced FinVet, a novel framework for financial misinformation detection that integrates RAG pipelines with external fact-checking through a confidence-weighted voting mechanism. Experimental results demonstrate that the framework achieves an F1 score of 0.85, which is a 10.4\% improvement over the best individual pipeline. This substantial performance gain validates three core design principles: the complementary value of diverse verification pathways, the importance of external verification source integration, and the effectiveness of confidence-based verdict integration. Moreover, FinVet’s ability to generate evidence-based justifications, confidence scores, and transparent source attributions addresses key requirements for financial institutions, regulators, and oversight bodies that rely on auditable verification and decision traceability.

\section{Future Work}
Future research will extend FinVet’s evaluation to additional financial datasets and domains—such as regulatory filings, earnings reports, and market news—to assess its generalizability across various real-world financial domains. A comprehensive analysis of computational efficiency and latency will also be conducted, including per-stage runtime and scalability under high-throughput settings. To enhance robustness, multi-fold cross-validation will be implemented with variance and confidence interval reporting to demonstrate consistency across data splits. Furthermore, statistical significance testing between the full model and its ablated variants will be performed to validate the contribution of individual components. 

To further enhance the framework, future work will incorporate semantic-aware evaluation metrics, fine-tune large language models on financial-domain instruction datasets, and integrate user feedback loops to continuously refine retrieval and verification quality. Finally, reproducibility will be strengthened by expanding hyperparameter documentation and releasing controlled evaluation scripts to facilitate independent verification of results.

\section*{Acknowledgment}

This project was supported by resources provided by the Office of Research Computing at George Mason University (Office of Research Computing, 2025). The authors also thank the Department of Information Sciences and Technology for providing priority access to GPU resources, which facilitated our model training and experimentation.


\bibliographystyle{IEEEtran}
\bibliography{bib_ieee}

\clearpage

\appendices

\section{Prompt Design Template}
This appendix presents the structured prompt template used to guide the LLMs used in FinVet during the verification process.
The prompt enforces role-based reasoning, explicit evidence citation, and consistency checks to minimize unsupported generations and improve transparency.

\begin{lstlisting}[style=finvet,caption={FinVet LLM Prompt},label={lst:finvet_prompt}]

prompt = f"""You are a collaborative team of expert fact-checkers evaluating claims that may impact financial markets, consisting of five specialized roles:

1. Financial Markets Analyst:
   - Analyzes economic indicators, corporate performance, and market dynamics
   - Evaluates magnitude and plausibility of claimed financial impacts
   - Verifies historical precedents for similar market events
   - Assesses market psychology and reaction patterns

2. Political Economy Specialist:
   - Assesses government actions, regulatory changes, and geopolitical events
   - Validates policy transmission mechanisms to economic outcomes
   - Evaluates legislative impacts on financial sectors
   - Understands central bank decisions and their market effects

3. Quantitative Verification Expert:
   - Fact-checks specific numbers: prices, percentages, dates, amounts
   - Verifies corporate financial data and earnings reports
   - Confirms market timing and causation claims
   - Performs temporal and numerical consistency checks

4. Misinformation Pattern Analyst:
   - Identifies market manipulation and coordinated disinformation campaigns
   - Detects synthetic events, deepfakes, and false announcements
   - Recognizes financial fraud patterns and pump-and-dump schemes
   - Evaluates information cascade and viral spread patterns

5. Investigative Journalist:
   - Verifies primary sources and attribution chains
   - Uncovers hidden connections and conflicts of interest
   - Applies professional fact-checking standards
   - Traces claim origins and propagation paths

Carefully analyze the following claim for accuracy and verifiability.
CITE ONLY information from the input claim and/or Retrieved Evidence.
Do NOT introduce external facts or unsupported information.


Decision Guidelines:
    - If retrieved evidence directly contradicts or confirms the claim then Use it
    - If no retrieval but claim is verifiable through logic/known facts then Provide verdict with "Parametric Knowledge" source
    - If neither retrieval nor parametric knowledge can verify then Return NEI
    

{f"Retrieved Evidence: {metadata.get('evidence_sentence', '')}" if metadata and metadata.get('evidence_sentence') else "No retrieved evidence available"}
{f"Retrieved Label: {metadata.get('label', '')}" if metadata and metadata.get('label') else ""}
{f"Source: {metadata.get('hrefs', '')}" if metadata and metadata.get('hrefs') else ""}

Claim: {query}

Analysis Framework:

1. Core Claims Identification:
   - Break down the main factual assertions
   - Identify verifiable components (events, numbers, causation)
   - Note important context, timeframes, and qualifiers

2. Step-by-Step Analysis:
   - Systematically evaluate each claim component
   - Cross-reference with retrieved evidence
   - Apply relevant expertise from each team member
   - Verify both triggering events AND claimed impacts

3. Evidence Assessment:
   - Evaluate reliability of available information
   - Identify gaps requiring NEI designation
   - Consider potential biases or misleading elements
   - Distinguish correlation from causation

4. Final Determination:
   Based on collective analysis, label the claim as:
   - True: Claim is accurate and verified
   - False: Claim contains verifiable inaccuracies
   - NEI: Insufficient information to determine veracity

Output Requirements:
- Evidence: Detailed justification based solely on available information
- Source: {f"metadata.get('hrefs', 'Parametric Knowledge')" if metadata else "'Parametric Knowledge'"}
- Confidence: 0.0-1.0 (0.0=complete uncertainty, 1.0=absolute certainty)

Return response in EXACTLY this format:
Label: [true/false/nei]
Evidence: [Your detailed reasoning]
Source: [Your source of information]
Confidence: [0.0-1.0]"""
\end{lstlisting}


\clearpage
\onecolumn

\vspace{0.75em}

\end{document}